\documentclass[11pt]{article}
\pdfoutput=1
\usepackage{amsmath,epsf,amssymb,latexsym,array,color}
\usepackage{graphicx,wasysym}
\usepackage{cancel}

\def\be{\begin{equation}}
\def\ee{\end{equation}}
\def\ba{\begin{array}}
\def\ea{\end{array}}
\newcommand{\beq}{\begin{equation}}
\newcommand{\eeq}[1]{\label{#1}\end{equation}}
\newcommand{\bea}{\begin{eqnarray}}
\newcommand{\eea}[1]{\label{#1}\end{eqnarray}}

\usepackage{ifpdf}
\ifx\pdfoutput\undefined
   \pdffalse
\else
   \pdfoutput=1
   \pdftrue
  \usepackage[pdftex]{hyperref}
\newtheorem{theorem}{Theorem}[section]

\begin{document}


\begin{titlepage}

\hskip 1.5cm

\begin{center}
{\huge \bf{Relaxed Supergravity}}
\vskip 0.8cm  
{\bf \large Hun Jang\footnote{hun.jang@nyu.edu}}  
\vskip 0.75cm

{\em Center for Cosmology and Particle Physics\\
	Department of Physics, New York University \\
	726 Broadway, New York, NY 10003, USA}
	
	\vspace{12pt}

\end{center}

\begin{abstract}
We propose a novel class of $\mathcal{N}=1$ supergravity (called ``Relaxed supergravity'') that can enlarge the space of scalar potentials, relaxing the strongly-constrained form of the prototype supergravity potential. It has very long been considered that in $\mathcal{N}=1$ supergravity a negative term of scalar potential can be given only by the gravitino-mass term (i.e. $-3e^{G}$) from the F-term potential, while such negative term is absent in global supersymmetry (SUSY). In this letter, however, we firstly discover a new negative-definite contribution to the scalar potential in $\mathcal{N}=1$ supergravity and even global SUSY. In the end, this allows us to have a general scalar potential. To achieve this, we start with detection of a ``no-go'' theorem for the higher order corrections in minimal supergravity models of inflation, which was investigated by Ferrara, Kallosh, Linde, and Porrati (FKLP). Based on the no-go theorem, we establish a superconformal action of a certain higher order correction for generating the new negative term. Then, we identify essential constraints on the new negative term by inspecting the suppression of the nonrenormalizable terms. We find that relaxed supergravity has a cutoff equal to SUSY breaking scale $M_S$. This signals that supersymmetry may be broken at high scale according to the cutoff leaving the naturalness issue aside. We suggest that relaxed supergravity can be a universal framework for building supergravity models of various phenomenologies under broken SUSY from particle physics to cosmology. 
\end{abstract}

\vskip 1 cm
\vspace{24pt}
\end{titlepage}
\tableofcontents

\section{Introduction}

The application of supergravity to inflationary cosmology has recently been of great interest and studied by many authors \cite{inf1,inf2,inf3,inf4,inf5,inf6}. However, it remains still challenging to build viable models of a certain phenomenology in the context of supergravity. For instance, it is not straightforward to realize both inflationary dynamics and minimal supersymmetric standard model (MSSM) at the same time in a unified setup. The first reason for the difficulty is due to the large hierarchy between Hubble scale $H\sim 10^{-5}M_{pl}$ for inflation and electroweak (TeV) scale of order $10^{-15}M_{pl}$ for the observable-sector dynamics of standard model (SM). The second is because {\it standard} supergravity predicts the complicated structure of the F-term scalar potential, i.e. $V_F=e^G(G_IG^{I\bar{J}}G_{\bar{J}}-3)$ where $G$ is the supergravity G-function defined by $G \equiv K + \ln W + \ln \bar{W}$ which consists of K\"{a}hler potential $K$ and superpotential $W$. This implies that one must always explore a proper choice of the supergravity G-function, which is unfortunately nontrivial in general. For these reasons, it is very demanding to construct phenomenologically-desirable scalar potentials within standard supergravity. 

The $\eta$ problem \cite{eta_problem} is an example of such a difficulty\footnote{This issue gives rise to a hardship for obtaining a very small slow-roll parameter $\eta$ such that $\eta \ll 1$ due to the exponentially growing behavior of the F-term potential.}. No-scale supergravity \cite{No_scale_SUGRA} may be a solution to the $\eta$ problem because the corresponding F-term potential can exactly vanish, i.e. $V_F =0$. This is established by a clever choice of the supergravity G-function. In fact, the gravitino mass term ``$-3e^G$'' plays a critical role in the no-scale cancellation. Interestingly, one can easily have such no-scale structure through a logarithmic K\"{a}hler potential of the volume modulus fields and constant superpotential in string theory \cite{No_scale_SUGRA,KKLT,KKLMMT}. However, certain choice of superpotential may spoil the ``exact'' cancellation of the F-term potential yielding a remnant as shown in Eq. (14) of Ref. \cite{FP}. Hence, no-scale supergravity is very sensitive to the given form of both K\"{a}hler potential and superpotential. Moreover, no-scale supergravity may cause a vast number of moduli, which correspond to degenerate vacua being along flat directions in scalar potential. This turns out that moduli stabilization, which is necessary to obtain a unique vacuum, is still required in no-scale supergravity. Thus, current no-scale supergravity is not a complete strategy for model building.

Recent developments in modification of the supergraivty scalar potential have been made, e.g. liberated supergravity recently proposed by Farakos, Kehagias, and Riotto \cite{Liberated} and various types of new Fayet-Iliopoulos (FI) terms proposed by many authors \cite{newFI1,newFI2,newFI3,newFI4,newFI5,newFI6}. In particular, liberated supergravity was the first attempt to allows us to have a general scalar potential. In fact, inspired by liberated supergravity, we investigate such a general scalar potential in the other fashion in this work. However, it has recently been found that liberated supergravity is not liberated literally due to strong constraints on the general function \cite{jp1,jp3}. On the contrary, new FI terms can modify only D-term potentials, which still have non-trivial field dependence and can give us only the non-negative-definite contribution to the scalar potential. Consequently, the recent studies do not have full generality of scalar potential.

Obviously, it has very long been thought of that a negative-definite term in scalar potential can be given only by the gravitino-mass term ``$-3e^G$'' in the standard $\mathcal{N}=1$ supergravity, and is not present in global supersymmetry (SUSY) \cite{fvp}. It is thus inevitable to acquire another type of cancellation in scalar potential through a new negative term so that we have a general scalar potential being beyond no-scale supergravity and the recent works. In that sense, it remains very intriguing to answer the following open questions: How can we obtain a new negative-definite potential term in supergravity? To what extent can we reform the supergravity scalar potential in a general fashion? We affirmatively answer these questions throughout this letter.

Our work is organized as follows. In Sec. 2, we revisit the higher order corrections in the minimal supergravity models of inflation constructed by Ferrara, Kallosh, Linde, and Porrati (FKLP) \cite{HOC}. We firstly identify a no-go theorem for the higher order corrections. The no-go theorem is supported by the fact that canonical kinetic term of the gauge field in the vector multiplet must be present in the supergravity lagrangian. In Sec. 3, we propose how to relax the strongly-constrained standard form of the scalar potential by adding a special choice of higher order correction to the standard supergravity. Here we discover a new negative-definite scalar potential as a general function. In addition, we find essential constraints on the new negative term by inspecting the suppression of nonrenormalizable lagrangians to ensure that our theory is self-consistent as an effective field theory. This leads to a cutoff which is identified with the high-scale SUSY breaking mass $M_S$ \cite{High_SUSY}. Next, using the new negative term, we present a relaxing procedure for generating a general scalar potential\footnote{In this sense, our proposal reserves the name ``Relaxed supergravity.''}. Plus, we compare our theory with the liberated supergravity by Farakos, Kehagias, and Riotto. Next, we briefly discuss the global SUSY limit of relaxed supergravity. Finally, in Sec. 4, we summarize our findings and give outlook on this work.

\section{No-go theorem for higher order corrections in FKLP model}

In this section, we revisit {\it higher order corrections} in the minimal supergravity models of inflation proposed by  Ferrara, Kallosh, Linde, and Porrati (FKLP) \cite{HOC}. First, we start with considering a vector multiplet $V$, its field strength multiplet $\bar{\lambda}P_L\lambda$, and real linear multiplet $(V)_D$ whose lowest component is given by the auxiliary field $D$ of the vector multiplet $V$ as follows:
\begin{eqnarray}
&& V = \{0,0,0,0,A_{\mu},\lambda,D\} ~\textrm{in the Wess-Zumino gauge,~i.e.}~v=\zeta=\mathcal{H}=0,\\
&& \bar{\lambda}P_L\lambda = (\bar{\lambda}P_L\lambda,-i\sqrt{2}P_L\Lambda,2D_-^2,0,+i\mathcal{D}_{\mu}(\bar{\lambda}P_L\lambda),0,0) = \{\bar{\lambda}P_L\lambda, P_L\Lambda,-D_-^2\},\\
&& \bar{\lambda}P_R\lambda = (\bar{\lambda}P_R\lambda,+i\sqrt{2}P_R\Lambda,0,2D_+^2,-i\mathcal{D}_{\mu}(\bar{\lambda}P_R\lambda),0,0) = \{\bar{\lambda}P_R\lambda, P_R\Lambda,-D_+^2\},\\
&& (V)_D = (D,\cancel{\mathcal{D}}\lambda,0,0,\mathcal{D}^{b}\hat{F}_{ab},-\cancel{\mathcal{D}}\cancel{\mathcal{D}}\lambda,-\square^CD),
\end{eqnarray}
where we have used the following notations\footnote{We follow the sign convention $(-,+,\cdots,+)$ for spacetime metric, and notations used in Ref. \cite{fvp}:\begin{eqnarray}
&& \mathcal{D}_{\mu}\lambda \equiv \bigg(\partial_{\mu}-\frac{3}{2}b_{\mu}+\frac{1}{4}w_{\mu}^{ab}\gamma_{ab}-\frac{3}{2}i\gamma_*\mathcal{A}_{\mu}\bigg)\lambda - \bigg(\frac{1}{4}\gamma^{ab}\hat{F}_{ab}+\frac{1}{2}i\gamma_* D\bigg)\psi_{\mu}\nonumber
\\
 && \hat{F}_{ab} \equiv F_{ab} + e_a^{~\mu}e_b^{~\nu} \bar{\psi}_{[\mu}\gamma_{\nu]}\lambda,\qquad F_{ab} \equiv e_a^{~\mu}e_b^{~\nu} (2\partial_{[\mu}A_{\nu]}),\nonumber\\
 && \hat{F}^{\pm}_{\mu\nu} \equiv \frac{1}{2}(\hat{F}_{\mu\nu}\pm \tilde{\hat{F}}_{\mu\nu}), \qquad \tilde{\hat{F}}_{\mu\nu} \equiv -\frac{1}{2} i\epsilon_{\mu\nu\rho\sigma}\hat{F}^{\rho\sigma}.\nonumber
\end{eqnarray}}
\begin{eqnarray}
&& P_L\Lambda \equiv \sqrt{2}P_L(-\frac{1}{2}\gamma\cdot \hat{F} + iD)\lambda,\qquad P_R\Lambda \equiv \sqrt{2}P_R(-\frac{1}{2}\gamma\cdot \hat{F} - iD)\lambda,\\
&& D_-^2 \equiv D^2 - \hat{F}^-\cdot\hat{F}^- - 2  \bar{\lambda}P_L\cancel{\mathcal{D}}\lambda,\qquad D_+^2 \equiv D^2 - \hat{F}^+\cdot\hat{F}^+ - 2  \bar{\lambda}P_R\cancel{\mathcal{D}}\lambda.
\end{eqnarray}
Then, after making the corrections to be generic as K\"{a}hler-invariant and field-dependent form, the higher order corrections to the standard supergravity action, which is given by Eq. (3.17) of Ref. \cite{HOC}, can be rewritten as 
\begin{eqnarray}
\mathcal{L}_{n} \supset \frac{(\bar{\lambda}P_L\lambda)^2(\bar{\lambda}P_R\lambda)^2}{(S_0\bar{S}_0e^{-K/3})^2}
T^k\Big( \frac{(\bar{\lambda}P_R\lambda)^2}{(S_0\bar{S}_0e^{-K/3})^2}\Big)\bar{T}^l\Big( \frac{(\bar{\lambda}P_L\lambda)^2}{(S_0\bar{S}_0e^{-K/3})^2}\Big)
\Big( \frac{(V)_D}{(S_0\bar{S}_0e^{-K/3})} \Big)^p\Psi_n(Z,\bar{Z})|_D, 
\end{eqnarray}
where $K(Z,\bar{Z})$ is a K\"{a}hler potential; $T$ is the chiral projection; $n=4+2k+2l+p$ with $n\geq 4$, and $\Psi_n(Z,\bar{Z})$ is a general real function of matter fields $Z$'s:
\begin{eqnarray}
&& Z^i = (z^i,-i\sqrt{2}P_L\chi^i,-2F^i,0,+i\mathcal{D}_{\mu}z^i,0,0) = \{ z^i, P_L\chi^i,F^i\},\\
&& \bar{Z}^{\bar{i}} = (\bar{z}^{\bar{i}},+i\sqrt{2}P_R\chi^{\bar{i}},0,-2\bar{F}^{\bar{i}},-i\mathcal{D}_{\mu}\bar{z}^{\bar{i}},0,0) = \{ \bar{z}^{\bar{i}}, P_R\chi^{\bar{i}},\bar{F}^{\bar{i}}\}.
\end{eqnarray}
We also use the superconformal compensator multiplet $S_0$: 
\begin{eqnarray}
&& S_0 = (s_0,-i\sqrt{2}P_L\chi^0,-2F_0,0,+i\mathcal{D}_{\mu}s_0,0,0) = \{ s_0, P_L\chi^0,F_0\},\\
&& \bar{S}_0 = (\bar{s}_0,+i\sqrt{2}P_R\chi^0,0,-2\bar{F}_0,-i\mathcal{D}_{\mu}\bar{s}_0,0,0) = \{\bar{s}_0, P_R\chi^0,\bar{F}_0\}.
\end{eqnarray}

Using the superconformal tensor calculus \cite{SuperconTensorCal,Linear}, we find the corresponding bosonic lagrangian as
\begin{eqnarray}
\mathcal{L}_n|_B \supset (F^{+2}-D^2 )^{1+k}(F^{-2}-D^2)^{1+l}D^p \Psi_n(z,\bar{z})|_D= \Big(\frac{F^2}{2}-D^2\Big)^{(n-p)/2}D^p\Psi_n(z,\bar{z})|_D,\label{highorder}
\end{eqnarray}
where $F^2 \equiv F_{\mu\nu}F^{\mu\nu}$ is the square of Maxwell tensor; $F^{\pm}_{\mu\nu} \equiv \frac{1}{2}(F_{\mu\nu} \pm \Tilde{F}_{\mu\nu})$, and $\Tilde{F}_{\mu\nu} \equiv -\frac{i}{2}\varepsilon_{\mu\nu\rho\sigma}F^{\rho\sigma}$ is the dual tensor. In the last line of Eq. \eqref{highorder}, we have used $F^{\pm2} = \frac{1}{2}F^2$ and $(n-p)/2 = 2+k+l$. We note that the lagrangian of many higher order corrections can be given by a polynomial of the terms with various powers of $n,p$. 

Now, let us consider the case when $p=0$. Then, defining $\hat{D}\equiv \frac{F^2}{2}-D^2$ and $\Psi_n\equiv \Psi_n(z,\bar{z})|_D$, we rewrite the bosonic lagrangian as 
\begin{eqnarray}
\mathcal{L}_{\textrm{higher order}}^{(p=0)}|_B = \sum_{n=4}^N \hat{D}^{n/2}\Psi_n = \hat{D}^2 \Psi_4 + \hat{D}^{5/2} \Psi_5+ \hat{D}^3 \Psi_6 + \cdots.
\end{eqnarray}
The standard supergravity is specified by the following superconformal action
\begin{eqnarray}
 \mathcal{L}_{standard} &=& -3[S_0\Bar{S}_0e^{-K(Z^A,\bar{Z}^{\bar{A}})/3}]_D + [S_0^3W(Z^A)]_F -\beta[ \bar{\lambda}P_L\lambda]_F + h.c.,
\end{eqnarray}
where we have used $\beta$ as a general normalization of the kinetic term of the vector field. The corresponding D-term lagrangian is then found to be
 \begin{eqnarray}
 \mathcal{L}_{standard}|_B = 2\beta D^2 - \beta (F^{+2}+F^{-2}) = -2\beta \Big(\frac{F^2}{2}-D^2\Big) \equiv -2\beta \hat{D}.
 \end{eqnarray}
 Taking both the standard and higher order terms, we find the general D-term lagrangian as
\begin{eqnarray}
\mathcal{L}_{tot}|_{B} = -2\beta \hat{D} + \hat{D}^2 \Psi_4 + \hat{D}^{5/2} \Psi_5+ \hat{D}^3 \Psi_6 + \cdots \equiv P(\hat{D}).
\end{eqnarray}
Notice that this lagrangian is a polynomial of $\hat{D}$. When solving the equation of motion for $D$, we gain 
\begin{eqnarray}
\frac{\partial \mathcal{L}_{tot}|_{B}}{\partial D} = \frac{\partial \hat{D}}{\partial D} \frac{\partial \mathcal{L}_{tot}|_{B}}{\partial \hat{D}} = 0 \quad \implies \quad D=0  \quad \textrm{ or } \quad \frac{\partial \mathcal{L}_{tot}|_{B}}{\partial \hat{D}}= \frac{\partial P(\hat{D})}{\partial \hat{D}}= 0.
\end{eqnarray}
If the trivial solution $D=0$ is unstable or supersymmetry is broken, we have to consider the non-vanishing solution for $D$. The non-trivial solution for $\hat{D}$ can be found by
\begin{eqnarray}
\hat{D} = \hat{D}(\Psi_4,\Psi_5,\Psi_6,\cdots).
\end{eqnarray}
Notice that there is no any dependence on Maxwell tensor term in the solution for $\hat{D}$! After integrating out $D$, we face the {\it unphysical} situation that the kinetic term of the vector field is always absent in the lagrangian for any $N$. We point out that this case must be physically excluded. So, we propose a no-go theorem for the higher order corrections in FKLP supergravity model of inflation as follows:
\begin{theorem}[No-go theorem for higher order corrections in FKLP model]~\\
An arbitrary combination of the standard term and higher order corrections without any power of the real linear multiplet $(V)_D$, i.e. $p$, cannot produce the gauge kinetic term, and thus must be excluded in a physical theory.\label{no_go_theorem}
\end{theorem}
Based on this no-go theorem for the higher order corrections, we speculate that one has to include some non-vanishing powers of $(V)_D$, i.e. $p$, in the higher order corrections in order to generate the correct kinetic term.

\section{Novel class of $\mathcal{N}=1$ supergravity: ``Relaxed supergravity''}

In this section, we propose a novel class of $\mathcal{N}=1$ supergravity, called ``Relaxed Supergravity,'' that enlarges the space of scalar potentials by considering the higher order correction in FKLP minimal supergravity models of inflation \cite{HOC}.

\subsection{Discovery of a new negative-definite term of scalar potential in supergravity}

For the vector multiplet $V$, as a setup, we suppose three ``NO'' things when constructing a superconformal action of supergravity containing some higher order corrections as follows:
\begin{itemize}
    \item {\bf No Fayet-Iliopoulos term.} There is no term linear in the auxiliary field $D$, i.e. no any Fayet-Iliopoulos $D$ term.
    \item {\bf No gauging.} The vector multiplet $V$ is {\it not} gauged.
    \item {\bf No-go theorem.} There must be some powers of the real linear multiplet $(V)_D$ in the higher order corrections in order to satisfy the no-go theorem \eqref{no_go_theorem}.
\end{itemize} These three assumptions will play a role in finding a new contribution to the scalar potential. Notice that the three conditions are {\it not applicable} for the other vector multiplets associated with conventional gauge groups.

Now, we are ready to consider a superconformal action of a certain higher order correction. We define 
\begin{eqnarray}
\mathcal{L}_{RS} \equiv \bigg[ -\frac{1}{4}(S_0\bar{S}_0 e^{-K/3})^{-4}  (\bar{\lambda}P_L\lambda)(\bar{\lambda}P_R\lambda) ((V)_D)^2 \frac{1}{\mathcal{U}(Z,\Bar{Z})^2}\bigg]_D,\label{RS}
\end{eqnarray}
where $S_0$ is the conformal compensator; $K(Z^I,\bar{Z}^{\bar{\imath}})$ is the supergravity K\"{a}hler potential of the matter chiral multiplets $Z^I$'s; $\bar{\lambda}P_L\lambda$ is the field strength multiplet corresponding to a vector multiplet $V$ whose fermionic superpartner is given by $\lambda$; $(V)_D$ is a real multiplet whose lowest component is given by the auxiliary field $D$ of the vector multiplet $V$, and $\mathcal{U}$ is defined as a general {\it gauge-invariant} real function of the matter multiplets. Therefore, including the standard supergravity terms, we reach the total superconformal action of our supergravity as
\begin{eqnarray}
 \mathcal{L} &=& -3[S_0\Bar{S}_0e^{-K(Z^A,\bar{Z}^{\bar{A}})/3}]_D + \bigg[ -\frac{1}{4}(S_0\bar{S}_0 e^{-K/3})^{-4}  (\bar{\lambda}P_L\lambda)(\bar{\lambda}P_R\lambda) ((V)_D)^2 \frac{1}{\mathcal{U}(Z,\Bar{Z})^2}\bigg]_D \nonumber\\
 && + [S_0^3W(Z^A)]_F -\frac{3}{4}[ \bar{\lambda}P_L\lambda]_F + h.c.
\end{eqnarray}
Notice that the numerical factor of the kinetic term for the vector multiplet $V$ is not $1/4$ but $3/4$. This different factor is set to yield the canonically normalized kinetic term of the vector field. The bosonic lagrangian of the auxiliary field $D$ is then found\footnote{The components of the superconformal action of relaxed supergravity will be spelled out in the forthcoming paper \cite{component_action_RS}.} by
\begin{eqnarray}
 \mathcal{L}_D &=& -\frac{3}{4} \Big(-D^2 + \frac{F^2}{2}-D^2 + \frac{F^2}{2}\Big) - \frac{1}{2\mathcal{U}^2}\Big(-D^2 + \frac{F^2}{2}\Big)\Big(-D^2 + \frac{F^2}{2}\Big)D^2 \nonumber\\
 &=& \frac{3}{2}D^2 -\frac{3}{4}F^2 +  \left( -\frac{D^6}{2\mathcal{U}^2} + \frac{D^4F^2}{2\mathcal{U}^2} -\frac{D^2F^4}{8\mathcal{U}^2}\right),
\end{eqnarray}
where $F^2 \equiv F_{\mu\nu}F^{\mu\nu}$. This gives the potential in $D$
\begin{eqnarray}
 V(D) = \frac{3}{4}F^2-\frac{3}{2}D^2 + \left( \frac{D^2F^4}{8\mathcal{U}^2} - \frac{D^4F^2}{2\mathcal{U}^2}+\frac{D^6}{2\mathcal{U}^2}\right).\label{Dpotential}
\end{eqnarray}

Here is a crucial remark. We should be careful about presence of the kinetic term for the auxiliary field $D$. Let us look at the composite superconformal multiplet $\mathcal{V}\equiv ((V)_D)^2$. Its highest component $\mathcal{D}_{\mathcal{V}}$ contains the second derivative term of the field $D$ with respect to the spacetime coordinates, i.e. $\mathcal{D}_{\mathcal{V}} \supset -2D\square^CD \sim (\partial D)^2 + \textrm{total derivative}$. We find that since $\mathcal{L}_{RS} \equiv -[\mathcal{R}\cdot \mathcal{V}]$ where $\mathcal{R}$ is defined to be the remaining parts except for $\mathcal{V}\equiv ((V)_D)^2$ and a minus sign, the relaxed supergravity action gives us a kinetic term of the field $D$ with the canonical sign, i.e. $\mathcal{L}_{RS} \supset -(\partial D)^2$ in the spacetime-metric convention $(-,+,\cdots,+)$. That is, $D$ is not a ghost but a physical field. From Eq.~\eqref{Dpotential}, we see that the canonically normalized field ``$\tilde{D}$'' such that $\Tilde{D} \equiv D/M_{pl}$ has a mass of Planck scale as follows: 
\begin{eqnarray}
\frac{\partial^2 V(D)}{\partial D^2}\bigg|_{D\sim \sqrt{\mathcal{U}}} = -3 + 15 \frac{D^4}{\mathcal{U}^2}\bigg|_{D\sim \sqrt{\mathcal{U}}}= -3 + 15 \equiv \frac{m_{\Tilde{D}}^2}{M_{pl}^2}.
\end{eqnarray}
 Accordingly, we are able to integrate out the $D$ degree of freedom with the Planck mass in the first place. 

After solving the equation of motion for $D$, we obtain the following solutions 
\begin{eqnarray}
 D= 0 , \qquad D^2 = \mathcal{U} \sqrt{1 + \frac{F^4}{36\mathcal{U}^2}} + \frac{F^2}{3},\label{Dsolution}
\end{eqnarray}
where the first corresponds to the supersymmetric case, while the latter corresponds to the non-supersymmetric case. Looking at the potential for $D$ in Eq.~\eqref{Dpotential}, we observe that the point at $D=0$ is unstable, and the vacua is located at $D \neq 0$. Therefore, in our model, supersymmetry is spontanesously broken like the Higgs mechanism.  

We point out that since $D^2>0$, the general function must be non-negative-definite, i.e. $\mathcal{U}>0$. Next, integrating out the field $D$, we obtain the bosonic lagrangian as 
\begin{eqnarray}
 \mathcal{L}_D = -\frac{1}{4}F^2 +  \mathcal{U} \sqrt{1 + \frac{F^4}{36\mathcal{U}^2}} + \frac{F^4}{36\mathcal{U}}\sqrt{1 + \frac{F^4}{36\mathcal{U}^2}} + \frac{F^6}{24\mathcal{U}^2}.\label{RS_Lagrangian}
\end{eqnarray}
Then, expanding Eq.~\eqref{RS_Lagrangian}, we have the following 
\begin{eqnarray}
 \mathcal{L}_D = -\frac{1}{4}F^2 + \mathcal{U} + \frac{F^4}{24\mathcal{U}} + \frac{F^6}{24\mathcal{U}^2} + \frac{F^8}{2592\mathcal{U}^3} + \textrm{higher order terms in}~ F^2.\label{RS_Lagrangian_expanded}
\end{eqnarray}
Notice that the lagrangian produces the correct kinetic term for the vector $V$, and a new negative contribution to the scalar potential
\begin{eqnarray}
 V_{RS} \equiv -\mathcal{U},
\end{eqnarray}
where $\mathcal{U}>0$. Hence, the total scalar potential can be written in general by
\begin{eqnarray}
 V_{tot} = V_D + V_F - \mathcal{U},\label{Total_General_Scalar_Potential}
\end{eqnarray}
where $V_D$ and $V_F$ are the standard D- and F-term potentials. Again, $\mathcal{U}$ is a positive generic function, so that $V_{RS}$ is a purely negative-definite.

\subsection{Constraints on the new negative term}

In this section, by inspecting the most singular nonrenormalizable lagrangians, we find constraints on $\mathcal{U}$. We use the same analysis already done in our previous study \cite{jp1,jp3}. We identify the most singular terms by checking four fermions (i.e. $\sim (\bar{\chi}P_L\chi)(\bar{\chi}P_R\chi)$) and derivative terms:
\begin{eqnarray}
 \mathcal{L}_F^{\textrm{on U}} &\supset&  \bigg\{ \frac{(\mathcal{U}^{(1)})^4}{\mathcal{U}^5}, ~ 
 \frac{(\mathcal{U}^{(1)})^2\mathcal{U}^{(2)}}{\mathcal{U}^5} ,~\frac{(\mathcal{U}^{(2)})^2}{\mathcal{U}^3}, 
 ~ \frac{\mathcal{U}^{(1)}\mathcal{U}^{(3)}}{\mathcal{U}^3}, ~
 \frac{\mathcal{U}^{(4)}}{\mathcal{U}^2}
 \bigg\} \mathcal{O}_F^{(d=12)}, \nonumber\\
 \mathcal{L}_F^{\textrm{on U}} &\supset&  \bigg\{ \frac{(\mathcal{U}^{(1)})^4}{\mathcal{U}^6}, ~ 
 \frac{(\mathcal{U}^{(1)})^2\mathcal{U}^{(2)}}{\mathcal{U}^6} ,~\frac{(\mathcal{U}^{(2)})^2}{\mathcal{U}^4}, 
 ~ \frac{\mathcal{U}^{(1)}\mathcal{U}^{(3)}}{\mathcal{U}^4}, ~
 \frac{\mathcal{U}^{(4)}}{\mathcal{U}^3}
 \bigg\} [\mathcal{O}_F^{(d=12)}F^2]^{(d=16)}, \nonumber\\
  \mathcal{L}_F^{\textrm{on K}} &\supset&  \bigg\{ \frac{(K^{(1)})^4}{\mathcal{U}M_{pl}^4}, ~ 
 \frac{(K^{(1)})^2K^{(2)}}{\mathcal{U}M_{pl}^2} ,~\frac{(K^{(2)})^2}{\mathcal{U}}, 
 ~ \frac{K^{(1)}K^{(3)}}{\mathcal{U}}, ~ 
 \frac{K^{(4)}}{\mathcal{U}}M_{pl}^2
 \bigg\} \mathcal{O}_F^{(d=12)}, \nonumber\\
 \mathcal{L}_F^{\textrm{on K}} &\supset&  \bigg\{ \frac{(K^{(1)})^4}{\mathcal{U}^2M_{pl}^8}, ~ 
 \frac{(K^{(1)})^2K^{(2)}}{\mathcal{U}^2M_{pl}^6} ,~\frac{(K^{(2)})^2}{\mathcal{U}^2M_{pl}^4}, 
 ~ \frac{K^{(1)}K^{(3)}}{\mathcal{U}^2M_{pl}^4}, ~ 
 \frac{K^{(4)}}{\mathcal{U}^2M_{pl}^2}
 \bigg\} [\mathcal{O}_F^{(d=12)}F^2]^{(d=16)}, \nonumber\\
 \mathcal{L}_F^{\textrm{on S}} &\supset& \frac{1}{\mathcal{U}}M_{pl}^{-4} \mathcal{O}_F^{(d=12)},\nonumber\\
 \mathcal{L}_F^{\textrm{on S}} &\supset& \frac{1}{\mathcal{U}^2}M_{pl}^{-4} [\mathcal{O}_F^{(d=12)}F^2]^{(d=16)},\nonumber\\
  \mathcal{L}_D &=& -\frac{1}{4}F^2 + \mathcal{U} + \frac{F^4}{24\mathcal{U}} + \frac{F^6}{24\mathcal{U}^2} + \frac{F^8}{2592\mathcal{U}^3} + \textrm{higher order terms in}~ F^2 \nonumber.
\end{eqnarray}
where $K$ is the K\"{a}hler potential; $\mathcal{U}$ is the general function; $\mathcal{O}_F^{(d=12)}$ only includes fermions, and $F^2 \equiv F_{\mu\nu}F^{\mu\nu}$. We denote $\mathcal{L}_F^{\textrm{on U/K}}$ by the lagrangians of the derivatives of $\mathcal{U}$ and $K$ with respect to the matter fields, while $\mathcal{L}_F^{\textrm{on S}}$ by those of the derivatives of $s_0\bar{s}_0$ with respect to the conformal compensator field. We observe that the strongest constraint comes from the D-term lagrangian 
\begin{eqnarray}
\mathcal{L}_D \supset \frac{F^4}{24\mathcal{U}} \implies \frac{1}{\mathcal{U}} \lesssim \frac{1}{\Lambda_{cut}^{4}} \implies \Lambda_{cut} \sim \mathcal{U}^{1/4} = M_S \lesssim M_{pl}, \label{main_constraint}
\end{eqnarray}
where the last inequality is given by the fact that $\Lambda_{cut} \lesssim M_{pl}$. This means that relaxed supergravity has a cutoff exactly at the SUSY breaking scale, and supersymmetry may be broken at high scale \cite{High_SUSY} according to the cutoff. Therefore, our model is basically an effective field theory with broken SUSY valid up to the low energies below the SUSY breaking scale $M_S$.

\subsection{Relaxation of scalar potential beyond no-scale supergravity}

In the previous section 3.2, we have seen that the total scalar potential in our theory is given by Eq. \eqref{Total_General_Scalar_Potential}
\begin{eqnarray}
V_{tot}(z^I,\bar{z}^{\bar{I}}) = V_D(z^I,\bar{z}^{\bar{I}}) + V_F(z^I,\bar{z}^{\bar{I}}) - \mathcal{U}(z^I,\bar{z}^{\bar{I}}),\nonumber
\end{eqnarray}
where the potentials are functions of matter fields $z^I$'s, and the new term is moderately constrained by Eq.~\eqref{main_constraint}. To analyze the new potential term, let us begin with a decomposition of matter multiplets as follows:
\begin{eqnarray}
Z^I \equiv (Z^s, Z^i),
\end{eqnarray}
where $Z^s$ is supposed to control the SUSY breaking scale $M_S$ in a hidden sector, while $Z^i$ are the normal matter ones that may belong to an observable sector. Next, we define
\begin{eqnarray}
\mathcal{U}(z^I,\bar{z}^{\bar{I}}) \equiv V_{\mathcal{U}}^{\cancel{S}}(z^I,\bar{z}^{\bar{I}}) - \sum_{a \neq \cancel{S}}V_{\mathcal{U}}^a(z^I,\bar{z}^{\bar{I}}) >0,\\
V_D(z^I,\bar{z}^{\bar{I}}) \equiv V_D^{\cancel{S}}(z^I,\bar{z}^{\bar{I}}) + \sum_{A\neq \cancel{S}} V_D^A(z^I,\bar{z}^{\bar{I}}) >0, 
\end{eqnarray}
where each potential $V_{\mathcal{U}}^a$ can be either negative or positive definite, and has a different energy scale such that $|V_{\mathcal{U}}^{\cancel{S}}| > |\sum_{a=1}V_{\mathcal{U}}^a|$. On the other hand, each D-term potential is positive semi-definite. Then, the total scalar potential is rewritten as
\begin{eqnarray}
V_{tot} =  V_D^{\cancel{S}}+ \Big( \sum_{A\neq \cancel{S}} V_D^A + V_F \Big) + \sum_{a}V_{\mathcal{U}}^a  - V_{\mathcal{U}}^{\cancel{S}}.
\end{eqnarray}
In order to have a maximally relaxed scalar potential, we may take the following choice
\begin{eqnarray}
V_{\mathcal{U}}^{\cancel{S}} \overset{!}{=}  V_D^{\cancel{S}} + \Big( \sum_{A\neq \cancel{S}} V_D^A + V_F \Big),\label{SUSY_scale_eq1}
\end{eqnarray}
which provides us the most general function form of the scalar potential
\begin{eqnarray}
V_{tot} = \sum_{a \neq \cancel{S}} V_{\mathcal{U}}^a(z^I,\bar{z}^I) < |V_{\mathcal{U}}^{\cancel{S}}|,
\end{eqnarray}
and the SUSY breaking scale $M_S$ such that
\begin{eqnarray}
M_S^4 = \mathcal{U} \equiv V_D^{\cancel{S}} + \Big( \sum_{A\neq \cancel{S}} V_D^A + V_F \Big) - V_{tot}.
\end{eqnarray}

Now, we have to explore under which conditions the general scalar potential can be well established. We may consider the following four suppositions:
\begin{itemize}
\item {\bf Partitioned gauge symmetries.} All $Z^i$'s must be neutral under any gauge group in which $Z^s$ is charged, and {\it vice versa}:
\begin{eqnarray}
&& G^{\cancel{S}}:~Z^s \rightarrow Z^s e^{q_s\Sigma} , \quad Z^i \rightarrow Z^i,\nonumber\\
&& G^i:~Z^s \rightarrow Z^s , \quad Z^i \rightarrow Z^i e^{q_i\Omega}
\end{eqnarray}
where $\Sigma$ and $\Omega$ are chiral multiplets as gauge parameters of the gauge groups $G^{\cancel{S}}$ and $G^i$, respectively. 
\item {\bf SUSY-breaking-scale cutoff dominance.} The scale of $V_D^{\cancel{S}}$ far exceeds the magnitude of any combination of the other potentials $V_D^A,V_F,V_{tot}$, so that the combination cannot cancel out $V_D^{\cancel{S}}$ and this solely controls the SUSY breaking scale $M_S$, i.e.
\begin{eqnarray}
|V_D^{\cancel{S}}| \gg \Big|\Big( \sum_{A\neq \cancel{S}} V_D^A + V_F \Big)-V_{tot}\Big| \implies \Lambda_{cut} = M_S = \mathcal{U}^{1/4} \approx |V_D^{\cancel{S}}|^{1/4} \neq0.\label{Dominance}
\end{eqnarray}
\item {\bf Broken supersymmetry.} We must have proper values of $z^s$ and $z^i$ such that $V_D^{\cancel{S}}\neq 0$ to protect broken SUSY all the times. 
\item {\bf Decomposition of scalar potential for moduli stabilization.} The total scalar potential must be decomposed into $z^s$-dependent and $z^s$-independent sectors in order to perform moduli stabilization for the fields $z^s$ in the simplest way, i.e.
\begin{eqnarray}
V_{tot} = V_{\mathcal{U}}^{s-depen}(z^s,\bar{z}^s) + V_{\mathcal{U}}^{s-indepen}(z^i,z^i).
\end{eqnarray}
If $V_{\mathcal{U}}^{s-depen}(z^s,\bar{z}^s)=0$, then we can choose any value of $z^s$ such that $V_D^{\cancel{S}}\neq0$, and $z^s$ becomes massless.
\end{itemize}
As long as the above conditions are satisfied, we are able to have the maximally relaxed scalar potential
\begin{eqnarray}
V_{tot} = \sum_{a \neq \cancel{S}}V_{\mathcal{U}}^a(z^I,\bar{z}^I)= V_{\mathcal{U}}^{s-depen}(z^s,\bar{z}^s) + \sum_{a\neq \cancel{S},s-depen}V_{\mathcal{U}}^a(z^i,\bar{z}^i) < |\underbrace{V_D^{\cancel{S}} + \Big( \sum_{A\neq \cancel{S}} V_D^A + V_F \Big)}_{=V_{\mathcal{U}}^{\cancel{S}}}|,
\end{eqnarray}
where the inequality comes from the condition $\mathcal{U}>0$.
 In the meantime, the corresponding constraint is given by
\begin{eqnarray}
\Lambda_{cut}^4=M_S^4= \underbrace{V_D^{\cancel{S}} + \Big( \sum_{A\neq \cancel{S}} V_D^A + V_F \Big)}_{=V_{\mathcal{U}}^{\cancel{S}}} - V_{tot}  \lesssim M_{pl}^4 \implies M_S^4 \sim V_D^{\cancel{S}}\lesssim M_{pl}^4,
\end{eqnarray}
where the last inequality is due to the dominance condition in Eq.~\eqref{Dominance}. Notice that $V_{D}^{\cancel{S}}$ is parametrically free up to the Planck scale $M_{pl}$, while the total scalar potential is parametrically free up to the SUSY breaking scale $M_S$.

Of course, one may wish to utilize the normal structures of the D- and F-term potentials in supergravity for some reasons. In this case, one can recover them by respecting the above assumptions in the following way: 
\begin{eqnarray}
&& V_{\mathcal{U}}^a(z^i) \supset V_F' \equiv A\cdot V_F(z^s=0)  = A\cdot e^G(G_IG^{I\bar{J}}G_{\bar{J}}-3) |_{z^s=0},\\
&& V_{\mathcal{U}}^a(z^i) \supset V_D' \equiv B\cdot V_D(z^s=0),
\end{eqnarray}
in which we have put $z^s=0$ in the usual D- and F-term potentials, and multiplied them by some arbitrary constants $A,B$ for generality. Thus, we have extra freedom in adjusting the scales of the D- and F-term potentials.

For example, the simplest toy model of relaxed supergravity can be given by the following. Let us consider an abelian $U_s(1)$ gauge symmetry. Assume that only a single matter field $z^s$ is charged under the $U_s(1)$, say $z^s \rightarrow e^{iq_s\theta}z^s$. Then, for a K\"{a}hler potential $K = -3\ln[T+\bar{T}-\frac{|z^s|^2+\delta_{i\bar{j}}z^i\bar{z}^{\bar{j}}}{3}]$ and a superpotential $W(T,z^i)$, a corresponding D-term potential is given by
\begin{eqnarray}
V_D^{\cancel{S}} = \frac{1}{2}g^2 q_s^2\frac{(z^s\bar{z}^s)^2}{(T+\bar{T})^2},\label{toy}
\end{eqnarray}
and the total scalar potential is given by
\begin{eqnarray}
V_{tot} = \underbrace{\sigma (|z^s|^2-\rho^2)^2}_{\textrm{s-dependent part}} + \underbrace{\sum_{a} V_{\mathcal{U}}^a(T,z^i)}_{\textrm{s-independent part}}  < V_D^{\cancel{S}}(z^s,T),\label{toy2}
\end{eqnarray}
where $\sigma,\rho$ are some constants, and we have used a potential $V_{\mathcal{U}}^a \supset \sigma (|z^s|^2-\rho^2)^2$ for producing a mass of $z^s$ in general. For this potential, we observe that $\left<z^s\right>=\rho \neq0$. Of course, it is straightforward for Eq.~\eqref{toy} to obey the dominance condition in Eq.~\eqref{Dominance} by choosing a large value of $\rho$. We see that SUSY breaking scale is determined by Eq.~\eqref{toy} while we have generic potentials in Eq.~\eqref{toy2}.

In this section, using a special choice in Eq.~\eqref{SUSY_scale_eq1}, we have treated a particular mechanism to derive a general scalar potential. However, there can be other mechanisms. These possibilities deserve further investigation in the future.

\subsection{``Relaxed'' supergravity versus ``Liberated'' supergravity}
Here we compare our relaxed supergraivity (RS) with the liberated supergravity (LS). First, let us recall the main result of the constraints on the liberated supergravity \cite{jp1}. The liberated term $\mathcal{U}$ that is added to the supergravity scalar potential as a general function of the matter fields is severely constrained by
\begin{eqnarray}
 \mathcal{U}^{(n)} \lesssim \begin{cases}
  \left(\dfrac{M_S}{M_{pl}}\right)^{8(4-n)}\left(\dfrac{M_{pl}}{\Lambda_{cut}}\right)^{2(4-n)} \quad \textrm{where}\quad0\leq n \leq 2\quad\textrm{for}~N_{mat} =1,\\
\left(\dfrac{M_S}{M_{pl}}\right)^{8(6-n)}\left(\dfrac{M_{pl}}{\Lambda_{cut}}\right)^{2(6-n)} \quad\textrm{where}\quad0\leq n \leq 4\quad\textrm{for}~N_{mat} \geq 2,
  \end{cases}\label{generic_constraints}
\end{eqnarray}
where $n$ is the order of the derivative with respect to the matter field, and $N_{mat}$ is the number of matter multiplets involved in a liberated supergravity theory of interest. The constraints correspond to the case when the matter fields are at their vacua. The scalar potential in the liberated supergravity must obey
\begin{eqnarray}
V_{LS} \equiv \mathcal{U} \lesssim \begin{cases}
  \left(\dfrac{M_S}{M_{pl}}\right)^{32}\left(\dfrac{M_{pl}}{\Lambda_{cut}}\right)^{8} \quad\textrm{for}~N_{mat} =1,\\
\left(\dfrac{M_S}{M_{pl}}\right)^{48}\left(\dfrac{M_{pl}}{\Lambda_{cut}}\right)^{12} \quad\textrm{for}~N_{mat} \geq 2,
  \end{cases}\label{generic_constraints}
\end{eqnarray}
On the other hand, in relaxed supergravity, we found that 
\begin{eqnarray}
V_{RS} = \sum_{a\neq \cancel{S}} V_{\mathcal{U}}^a < V_{\mathcal{U}}^{\cancel{S}} \sim V_D^{\cancel{S}} \sim M_S^4 \lesssim M_{pl}^4.
\end{eqnarray}
For instance, when we consider $\Lambda_{cut} = 10^{-2} M_{pl}$, we obtain
\begin{eqnarray}
 V_{RS} < 10^{-8}M_{pl}^4, \quad V_{LS} \lesssim 10^{-64}M_{pl}^4 \quad  \textrm{for}\quad N_{mat}=1, \quad  V_{LS} \lesssim 10^{-96}M_{pl}^4 \quad  \textrm{for}\quad N_{mat}\geq 2.
\end{eqnarray} 
Note that $V_{RS}$ can describe the inflation scale $\mathcal{O}(H^2M_{pl}^2)$ since it is bounded by parametrically free $V_{\mathcal{U}}^{\cancel{S}}$ up to the Planck scale $M_{pl}$, while any of $V_{LS}$'s cannot. This shows that relaxed supergravity excels the liberated one in defining a scalar potential at a desired energy level.  

\subsection{The first negative term of scalar potential in global supersymmetry}

We briefly discuss an intriguing physical implication on our findings in relaxed supergravity. It is well known that when supergravity is turned off (i.e. $M_{pl} \rightarrow \infty$), the scalar potential of the standard supergravity reduces to that of global SUSY. This is because in the limit we have $M_{pl}^4e^{G} \rightarrow 0$ and $M_{pl}^4e^G(G_IG^{I\bar{J}}G_{\bar{J}}) \rightarrow W_IK^{I\bar{J}}W_{\bar{J}}$ in the F-term potential $V_F$ where $G \equiv \frac{K}{M_{pl}^2} + \ln \frac{W}{M_{pl}^3}+ \ln \frac{\bar{W}}{M_{pl}^3}$ after recovering the Planck mass dimension $M_{pl}$. In particular, the relaxing term $\mathcal{U}$ can be alive in global SUSY since the bosonic lagrangians in Eqs.~\eqref{RS_Lagrangian} and \eqref{RS_Lagrangian_expanded} are independent of Planck mass $M_{pl}$. Of course, the general function $\mathcal{U}$ changes the total scalar potential in the same way as follows:
\begin{eqnarray}
V_{tot} = \underbrace{\frac{1}{2}|D^a|^2 + |W_I|^2}_{\textrm{standard global SUSY}} -~ \mathcal{U},\label{New_Global_SUSY}
\end{eqnarray}
where $D^a$ is the D-term solution with respect to some gauge killing vector fields $k^a(z)$, and $W_I \equiv \partial W/\partial z^I$ is the field derivative of superpotential. The result in Eq.~\eqref{New_Global_SUSY} is surprising in that it gives us the first {\it negative} contribution to the scalar potential in global supersymmery, allowing us to have any of Minkowski and (Anti) de Sitter spacetimes. Surely, it has long been regarded that there is only positive potentials in global SUSY, and thus either Minkowski or de Sitter spacetime is possible to exist. We expect that this new aspect may alter some known arguments led by the fact that the global SUSY scalar potential is always semi-positive, i.e. $V_{tot} = \frac{1}{2}|D^a|^2 + |F^I|^2 \geq 0$. We do not explore this here since it is beyond the scope of our purpose in this letter.

\section{Conclusion and outlook}

We summarize our findings and discuss some outlooks on our proposal. First, we have found a no-go theorem for the higher order corrections in FKLP minimal supergravity models of inflation. The no-go theorem tells us that the higher order corrections in the field strength of a vector multiplet $V$ must include some powers of the real linear multiplet $(V)_D$ whose lowest component is given by the auxiliary field $D$ of the vector multiplet $V$. In this work, we considered $((V)_D)^2$ as shown in Eq. \eqref{RS} to introduce a superconformal action of the higher order correction. Next, from the action, we discovered a new negative-definite generic potential term. We also found moderate constraints on the new negative term by evaluating the suppression of nonrenormalizable interactions with respect to a cutoff scale $\Lambda_{cut}$. These constraints identify the cutoff $\Lambda_{cut}$ with the SUSY breaking scale $M_S$. It turns out that we have to take into account high-scale SUSY breaking putting the naturalness away. Then, we showed a comparison between relaxed and liberated supergravities. As an example, we observed that we can generate the inflation energy of order $10^{-10}M_{pl}^4$ through relaxed supergravity since the total scalar potential is bounded above by $V_{D}^{\cancel{S}}$ less than the Planck scale $M_{pl}$, while we cannot do through the liberated supergravity because only the scales below $10^{-64}M_{pl}^4$ or $10^{-96}M_{pl}^4$ can be allowed. In this sense, relaxed supergravity is truly liberated than the original liberated supergravity. Furthermore, we have seen that even in the limit for global SUSY (i.e. $M_{pl} \rightarrow \infty$) the relaxing term ``$-\mathcal{U}$'' of relaxed supergravity can be present as a negative contribution to the scalar potential unlike the standard globally supersymmetric theory. 

Especially, we have presented a relaxing procedure of the scalar potential by requiring four conditions. The first is that $z^s$ ($z^i$) is charged but $z^i$ ($z^s$) is neutral under a gauge group $G^{\cancel{S}}$ ($G^i$). The second is that the scale of $V_{D}^{\cancel{S}}$ governing the SUSY-breaking scale $M_S$ rather exceeds those of the other potentials $V_D^A,V_F,V_{tot}$ satisfying Eq.~\eqref{Dominance}. The third is that values of $z^s$ and $z^i$ must hold non-vanishing $V_{D}^{\cancel{S}}$. The last is that the total scalar potential is decomposed into $z^s$-dependent and independent sectors to do moduli stabilization for the fields $z^s$ in the simplest way.  

Lastly, we discuss outlook on relaxed supergravity. First, one may wish to explain some phenomenologies from particle physics to cosmology in the context of either locally or globally supersymmetric theory. We suggest that our proposal can be utilized for constructing both supergravity and globally supersymmetric models of particle and cosmological phenomenologies. This is based on the fact that our supergravity predicts a general scalar potential up to the Planck energy $M_{pl}$, and the relaxing term can emerge in both theories in a consistent fashion. Second, we remark that the string realization of the superconformal action of the relaxing term deserves future investigation like the work of Ref. \cite{StringRealization}. Third, it would be worth studying to explore if other relaxing mechanisms can possibly exist in different setups beyond this work. Fourth, since our model has a cutoff $\Lambda_{cut}$ equal to the SUSY breaking scale $M_S$, one may study improved versions of relaxed supergravity which has a sufficiently large hierarchy between cutoff and SUSY breaking scale in order to recover the naturalness in the future. The last is that one may explore physical implications which are deduced by the first negative scalar potential in global SUSY.


\subsection*{Acknowledgments} 
H.J.\ is deeply grateful to Massimo Porrati for helpful discussions and comments, and thanks Osmin Lacombe for useful discussion. H.J.\ is fully supported by James Arthur Graduate Associate (JAGA) Fellowship from the Center for Cosmology and Particle Physics (CCPP) in Department of Physics at New York University.


\begin{thebibliography}{99}
\bibitem{component_action_RS}
H.~Jang, ``Component action of relaxed supergravity in superconformal formalism,'' to be forthcoming.
\end{thebibliography}
\end{document}